\documentclass[twocolumn,showpacs,preprintnumbers,amsmath,amssymb]{revtex4}
\usepackage{mathrsfs}
\usepackage{amsfonts}

\usepackage{graphicx}
\usepackage{}% Include figure files
\usepackage{amsmath}% Include figure files
\usepackage{dcolumn}% Align table columns on decimal point
\usepackage{color}

\begin{document}
\title{Control of microwave signals using bichromatic electromechanically induced transparency in three-mode circuit electromechanical systems}
\author{Cheng Jiang}
\email{chengjiang8402@163.com}
\author{Yuanshun Cui}
\author{Xiaowei Li}
 \affiliation{School of Physics and Electronic Electrical Engineering, Huaiyin Normal University, 111 West Chang Jiang Road, Huaian 223001, China}
\date{\today}
\begin{abstract}
We theoretically investigate the tunable slowing and advancing of microwave signals based on bichromatic electromechanically induced transparency in a three-mode circuit electromechanical system, where two mechanical oscillators with closely spaced frequencies are independently coupled to a common microwave cavity. In the presence of a strong microwave pump field, we obtain two transparency windows accompanied by steep phase dispersion in the transmitted microwave probe field. The width of the transparency window and the group delay of the probe field can be controlled effectively by the power of the pump field. It is shown that the maximum group delay of 0.12 ms and the advance of 0.27 ms can be obtained in the current experiments.
\end{abstract}

\maketitle
 \section{Introduction}
The field of cavity optomechanics and electromechanics \cite{Kippenberg2,Marquardt,Aspelmeyer,Aspelmeyer2}, which explores the radiation pressure interaction between mechanical and electromagnetic degrees of freedom, has witnessed remarkable progresses in the past decade. Cavity opto- and electromechanical systems allow us to cool the mechanical oscillators to their quantum ground state \cite{Teufel2,Chan}, observe the quantum zero-point motion \cite{Safavi-Naeini,Brahms}, and create squeezed light \cite{Purdy,Naeini2}. Moreover, the electromagnetic response of these systems is modified due to mechanical interactions, leading to the phenomena of normal-mode splitting \cite{Groblacher}, opto-/electromechanically induced transparency (OMIT/EMIT) \cite{Agarwal,Weis,Naeini,Karuza,Liu,Kronwald}, electromechanically induced absorption (EMIA) \cite{Hocke}, and microwave amplification \cite{Massel}. OMIT is the analog of electromagnetically induced transparency (EIT) \cite{Fleischhauer}, where an opaque medium can become transparent to the weak probe field when applying a strong pump beam. EIT has been first observed in atomic vapors and recently in various solid state systems such as quantum wells, dots, and nitrogen-vacancy centers \cite{Phillips,Santori,Xu}. It has been shown to be important in a variety of applications such as slow light \cite{Hau}, light storage \cite{Liu2}, and enhancement of nonlinear process \cite{Harris}, and so on. Likewise, OMIT, accompanied by steep phase dispersion, has also been used to control the group delay of the probe field in optomechanical systems both in optical \cite{Naeini,Tarhan} and microwave domains \cite{Jiang,Zhou}.

Recently, multimode optomechanical systems comprising more than two active degrees of freedom have gained increased attention \cite{Heinrich,Seok,Massel2,Wang}. One important class of multimode optomechanical systems consist of devices where a single electromagnetic mode couples to multiple mechanical modes. In the three-mode optomechanical systems where two distinct mechanical oscillators are coupled to a single electromagnetic cavity, phenomena of hybridization \cite{Massel2,Shkarin} and synchronization \cite{Zhang,Bagheri} of mechanical modes, two-mode back-action-evading measurements \cite{Woolley1}, and two-mode squeezed states \cite{Woolley2} have been investigated. Moreover, Shahidani \emph{et al.} \cite{Shahidani} have theoretically shown how to control and manipulate EIT in an optomechanical system with two moving mirrors which contains a Kerr-down-conversion nonlinear crystal. Wang \emph{et al.} \cite{Wang2} have recently studied the optomechanical analog of two-color EIT in a hybrid optomechanical system consisting of a cavity and a mechanical resonator with a two-level system (qubit). Two-color EIT has already been widely discussed in a variety of atomic systems \cite{Yang,Han}.

In this paper, we first investigate the microwave response of the three-mode circuit electromechanical system realized in Ref. \cite{Massel2} to a weak probe field in the presence of a strong pump field. We find that there are two transparency windows in the probe transmission spectrum, which can be named as bichromatic electromechanically induced transparency (EMIT). Using the experimental parameters, then we numerically calculate the group delay of the microwave probe field as a function of the pump power. Our results show that the maximum group delay of the transmitted probe field is about 0.12 ms and the signal advance of the reflected probe field is close to 0.27 ms, which can be further increased by reducing the damping rate of the mechanical oscillator. This hybrid system with bichromatic EMIT provides a good medium for controlling microwave photons at different frequencies and may find applications in quantum information processing.

The paper is organized as follows. In section \uppercase\expandafter{\romannumeral2}, we  describe the theoretical model and the equations of motion for the system operators. The expressions for the transmission and group delay of the probe field are given according to the input-output theory. In Sec. \uppercase\expandafter{\romannumeral3}, we present the numerical results and discuss the slowing and advancing of microwave signals based on bichromatic EMIT. We finally conclude the paper in Sec. \uppercase\expandafter{\romannumeral4}.

\begin{figure}
\includegraphics[width=7.5cm]{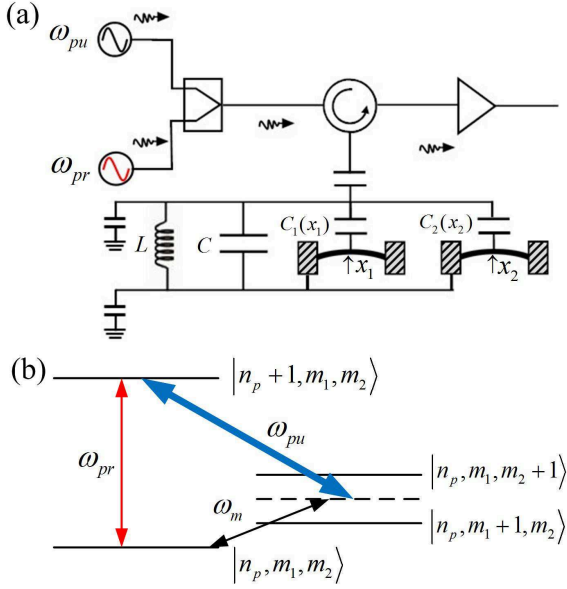}
\caption{(a) Schematic of a three-mode circuit electromechanical system in which two mechanical oscillators are coupled to a single microwave cavity denoted by equivalent inductance $L$ and equivalent capacitance $C$. The mechanical displacements $x_1$ and $x_2$ independently modulates the total capacitance $C$, and hence, the cavity frequency $\omega_c$. A strong pump field at the frequency $\omega_{pu}$ and a weak probe field at the frequency are applied to the microwave cavity simultaneously. (b) Level diagram of the hybrid system when the pump field is red tuned by an amount $\omega_m=(\omega_1+\omega_2)/2$. }
\end{figure}
\section{Model and Theory}
We consider a three-mode circuit electromechanical system as shown in Fig. 1, where two mechanical oscillators $b_k$($k=1, 2$) are parametrically coupled to a common microwave cavity $a$. The motion of each mechanical osciallator, expressed as time-varying capacitors $C_k$, independently modulates the total capacitance $C$, and hence, the cavity frequency $\omega_c$. The single-photon coupling strength $g_k=G_k x_{zp,k}$ is the product of $G_k=(\omega_c/2C)\partial C_k/\partial x_k$, denoting the cavity resonance frequency change on each mechanical displacement, and the mechanical osciallor's zero-point fluctuation $x_{zp,k}=\sqrt{\hbar/2m_k\omega_k}$ ($m_k$ and $\omega_k$ are the effective mass and resonance frequency of each oscillator, respectively). The microwave cavity is driven by a strong pump filed $E_{pu}$ with frequency $\omega_{pu}$ and a weak probe field $E_{pr}$ with frequency $\omega_{pr}$ simultaneously.
 In a rotating frame at the pump frequency $\omega_{pu}$, the Hamiltonian of the three-mode electromechanical system reads as
follows \cite{Massel2}:
\begin{eqnarray}
H=&&\hbar\Delta_{pu}a^\dagger a+\sum_{k=1,2}\hbar\omega_{k}b_{k}^\dagger b_{k}-\sum_{k=1,2}\hbar g_k
a^\dagger a(b_{k}^\dagger+b)\nonumber\\&&+i\hbar\sqrt{\kappa_{e}/2}E_{pu}(a^\dagger-a)\nonumber\\&&+i\hbar\sqrt{\kappa_{e}/2}E_{pr}(a^\dagger e^{-i\Omega t}-a e^{i\Omega t}).
\end{eqnarray}
The first term describes the energy of the microwave cavity mode, where
$a^\dagger$ $(a)$ is the creation (annihilation) operator of the cavity mode and $\Delta_{pu}=\omega_c-\omega_{pu}$ is the cavity-pump
field detuning. The second term gives the energy of the two
mechanical modes with creation (annihilation) operator
$b_k^\dagger$ $(b_k)$. The third term denotes the interaction between the cavity field and the two mechanical oscillators. The last two terms represent the interaction between the cavity field and the two input fields, where $E_{pu}$ and $E_{pr}$ are related to the power of the applied microwave fields by $\left\vert E_{pu}\right\vert=\sqrt{2P_{pu}/\hbar\omega_{pu}}$ and $\left\vert E_{pr}\right\vert=\sqrt{2P_{pr}/\hbar\omega_{pr}}$, respectively. $\Omega=\omega_{pr}-\omega_{pu}$ is the detuning between the probe filed and the pump field.
The microwave cavity has a linewidth of $\kappa=\kappa_{i}+\kappa_{e}$, where $\kappa_{i}$ is the intrinsic decay rate and $\kappa_{e}=\eta_c\kappa$ is due to external coupling to the feedline.

The dynamics of the system is described by a set of nonlinear Langevin equations. Starting from the Hamiltonian and applying the Heisenberg equations of motion for operators $a$ and $Q_k$ ($k=1,2$) [which is defined as dimensionless amplitude of the mechanical oscillations $Q_k=b_k^\dagger+b_k$], we derive the following quantum Langevin equations:
\begin{eqnarray}
\dot{a}=-\left[i(\Delta_{pu}-g_1Q_1-g_2Q_2)+\kappa/2\right] a\nonumber\\+\sqrt{\kappa_{e}/2}(E_{pu}+E_{pr}e^{-i\Omega t})+\sqrt{\kappa_e}a_{in},\\
\ddot{Q_1}+\gamma_1\dot{Q_1}+\omega_1^2 Q_1=2g_1\omega_1a^\dagger a+\xi_1,\\
\ddot{Q_2}+\gamma_2\dot{Q_2}+\omega_2^2 Q_2=2g_2\omega_2a^\dagger a+\xi_2,
\end{eqnarray}
where the decay rates for the microwave cavity ($\kappa$) and the two mechanical osicllators ($\gamma_1, \gamma_2$) have been introduced classically, and $a_{in}$ and $\xi_1 (\xi_2)$ are the zero-mean quantum and thermal noise terms. Under the assumption that the pump field is much stronger than the probe field ($E_{pu}\gg E_{pr}$), we derive the steady-state solutions to Eqs. (2)-(4) by setting all the time derivatives to zero. They are given by
\begin{eqnarray}
a_{s}=\frac{\sqrt{\kappa_{e}/2}E_{pu}}{\kappa/2+i\Delta},Q_{s,1}=\frac{2g_1\left\vert a_{s}\right\vert^2}{\omega_1},
Q_{s,2}=\frac{2g_2\left\vert a_{s}\right\vert^2}{\omega_2},
\end{eqnarray}
where $\Delta=\Delta_{pu}-g_1 Q_{s,1}-g_2 Q_{s,2}$ is the effective cavity detuning including radiation pressure effects. Without loss of generality, we have assumed that the steady-state value $a_{s}$ to be real and positive. Subsequently, we follow the typical procedure and solve Eqs. (2)-(4) perturbatively by rewriting each Heisenberg operator as the sum of its steady-state mean value and a small fluctuation with zero mean value,
\begin{equation}
a=a_{s}+\delta a, Q_1=Q_{s,1}+\delta Q_1, Q_2=Q_{s,2}+\delta Q_2.
\end{equation}
Inserting these equations into the Langevin equations Eqs. (2)-(4) and assuming $\left\vert a_{s}\right\vert\gg1$,
one can safely neglect the second-order small terms such as $\delta a^\dagger\delta a$, $\delta a\delta Q_1$, and $\delta a\delta Q_2$ which can result in higher-order sideband generation \cite{Xiong1}. In addition, since the drives are weak, but classical coherent fields, we will identify all operators with their
expectation values, and drop the quantum and thermal noise terms
\cite{Weis}. Then the linearized quantum Langevin equations can be obtained as follows:
\begin{eqnarray}
\frac{d}{dt}\delta a=&-(\kappa/2+i\Delta)\delta a+i(g_1\delta Q_1+g_2\delta Q_2)a_{s}\nonumber\\&+\sqrt{\kappa_{e}/2}E_{pr}e^{-i\Omega t},
\end{eqnarray}
\begin{equation}
\frac{d^2}{dt^2}\delta Q_1+\gamma_1\frac{d}{dt}\delta Q_1+\omega_1^2\delta Q_1=2\omega_1 g_1(a_{s}\delta a^\dagger+a_{s}^*\delta a),
\end{equation}
\begin{equation}
\frac{d^2}{dt^2}\delta Q_2+\gamma_2\frac{d}{dt}\delta Q_2+\omega_2^2\delta Q_2=2\omega_2 g_2(a_{s}\delta a^\dagger+a_{s}^*\delta a).
\end{equation}
In order to solve the linearized equations (7)-(9), we
use the ansatz \cite{Boyd} $\delta a(t)=a_{+}e^{-i\Omega
t}+a_{-}e^{i\Omega t}$, $\delta Q_1(t)=Q_{+,1}e^{-i\Omega
t}+Q_{-,1}e^{i\Omega t}$, and $\delta Q_2(t)=Q_{+,2}e^{-i\Omega
t}+Q_{-,2}e^{i\Omega t}$. Upon substituting the above ansatz into Eqs. (7)-(9), we derive the following solution:
\begin{eqnarray}
a_{+}=\frac{\kappa/2-i\delta-i\Delta_{pu}+\theta}{(\kappa/2-i\delta)^2+(\Delta_{pu}+i\theta)^2-\beta}\sqrt{\kappa_{e}/2}E_{pr},
\end{eqnarray}
where
\begin{eqnarray}
&&\alpha_1=\frac{g_1^2}{\omega_1^2},\alpha_2=\frac{g_2^2}{\omega_2^2},\nonumber\\&&\eta_1=\frac{\omega_1^2}{\omega_1^2-i\gamma_1\delta-\delta^2},
\eta_2=\frac{\omega_2^2}{\omega_2^2-i\gamma_2\delta-\delta^2},\nonumber\\&&
\beta=4n_p^2(\alpha_1\omega_1\eta_1+\alpha_2\omega_2\eta_2)^2,\nonumber\\&&\theta=2i n_p[\alpha_1\omega_1(\eta_1+1)+\alpha_2\omega_2(\eta_2+1)],
\end{eqnarray}
and $n_{p}=\left\vert a_{s}\right\vert^2.$
Here $n_{p}$, approximately equal to the number of
pump photons in the cavity, is determined by the following equation:
\begin{eqnarray}
n_p[(\frac{\kappa}{2})^2+(\Delta_{pu}-\frac{2g_1^2n_p}{\omega_1}-\frac{2g_2^2n_p}{\omega_2})^2]=\frac{\kappa_{e}}{2}E_{pu}^2.
\end{eqnarray}
The microwave response of the three-mode electromechanical system to the probe field under the influence of the strong pump field can be obtained by calculating the output field transmitted by the cavity. According to the input-output relation \cite{Gardiner}
$a_{out}(t)=a_{in}(t)-\sqrt{\kappa_e/2}a(t)$, one can obtain the output field
\begin{eqnarray}
a_{out}(t)=&&(E_{pu}-\sqrt{\kappa_{e}/2}a_{s})e^{-i\omega_{pu}t}\nonumber\\&&+(E_{pr}-\sqrt{\kappa_{e}/2}a_{+})
e^{-i(\Omega+\omega_{pu})t}\nonumber\\&&-\sqrt{\kappa_{e}/2}a_{-}e^{i(\Omega-\omega_{pu})t}.
\end{eqnarray}
We can see from Eq. (13) that the output field contains two input components ($\omega_{pu}$ and $\omega_{pr}$) and one generated four-wave mixing (FWM) component at the frequency $2\omega_{pu}-\omega_{pr}$. The transmission of the probe field, defined by the ratio of the
output and input field amplitudes at the probe frequency, is then
given by:
\begin{eqnarray}
t_p&=&\frac{E_{pr}-\sqrt{\kappa_{e}}a_{+}}{E_{pr}}\nonumber\\&=&1-\frac{\kappa_{e}}{2}\frac{\kappa/2-i\delta-i\Delta_{pu}+\theta}
{(\kappa/2-i\delta)^2+(\Delta_{pu}+i\theta)^2-\beta}.
\end{eqnarray}
The pump filed can not only modify the transmission of the probe filed, but also results in a rapid phase dispersion $\phi=\arg(t_p)$ of the transmitted probe field across the transmission window, which can lead to significant group delay expressed as
\begin{eqnarray}
\tau_{g}=\frac{\partial\phi}{\partial\Omega}.
\end{eqnarray}

\begin{figure}
\includegraphics[width=8cm]{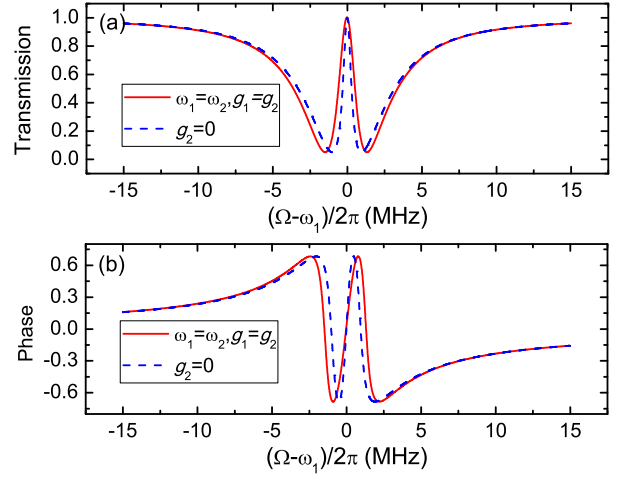}
\caption{(a) The magnitude and (b) phase of the probe transmission as a function of the $(\Omega-\omega_1)/2\pi$ in the two cases, where the two mechanical oscillators have the same frequency ($\omega_1=\omega_2$) and there is only one mechanical oscillator ($g_2=0$). The pump field is tuned to the red sideband of the cavity resonance ($\Delta_{pu}=\omega_1$) with $P_{pu}=4$ $\mu$W.
Other parameters used are $\omega_c=2\pi\times6.986$ GHz, $\kappa=2\pi\times6.2$ MHz, $\kappa_{e}=2\pi\times4.8$ MHz, $\omega_1=2\pi\times32.1$ MHz, $\gamma_1=\gamma_2=\gamma_m=2\pi\times930$ Hz, $g_1=2\pi\times39$ Hz.}
\end{figure}
\section{Numerical results and discussion}
In our numerical calcultions, we use the parameters of a realistic three-mode circuit electromechanical
system \cite{Massel2}: $\omega_c=2\pi\times6.986$ GHz, $\kappa=2\pi\times6.2$ MHz, $\kappa_{e}=2\pi\times4.8$ MHz, $\omega_1=2\pi\times32.1$ MHz, $\omega_2=2\pi\times32.5$ MHz, $\gamma_1=\gamma_2=\gamma_m=2\pi\times930$ Hz, $g_1=2\pi\times39$ Hz, $g_2=2\pi\times44$ Hz. We can see that
$\omega_{1}>\kappa$ and $\omega_2>\kappa$, therefore the system operates in the resolved-sideband regime, which is beneficial for the electromechanically induced transparency. In what follows, we investigate the microwave response of this coupled system in the two cases of equal and different mechanical frequencies separately.

First, we consider the case in which the resonance frequencies of the two mechanical oscillators are the same, i.e., $\omega_1=\omega_2=2\pi\times32.1$ MHz. In addition, their optomechanical coupling strengths are equal, we assume $g_1=g_2=2\pi\times39$ Hz. Figure 2 plots the transmission and phase of the transmitted probe field as a function of the probe detuning $(\Omega-\omega_1)/2\pi$ when the pump field is tuned to the red sideband of the cavity resonance $\Delta_{pu}=\omega_1$ with $P_{pu}=4$ $\mu$W. The red solid curve corresponds to the situation where $\omega_1=\omega_2$ and the blue dashed curve represents the case in the absence of the second mechanical oscillator. In both cases, it can be seen from Fig. 2(a) that a single transparency window appears when the beat frequency $\Omega$ is nearly resonant with the frequency of the mechanical oscillator, which is due to the destructive interference between the probe field and the generated anti-Stokes field. Such a phenomenon, usually termed as optomechanically induced transparency (OMIT) or electromechanically induced transparency (EMIT), has been well understood in the generic optomechanical systems where a mechanical resonator is coupled to a electromagnetic cavity \cite{Weis,Teufel1}. The concomitant phase dispersion in Fig. 2(b) indicates that this effect can be used to slow and advance electromagnetic fields in mechanical degrees of freedom \cite{Naeini,Zhou}.\begin{figure}
\includegraphics[width=8cm]{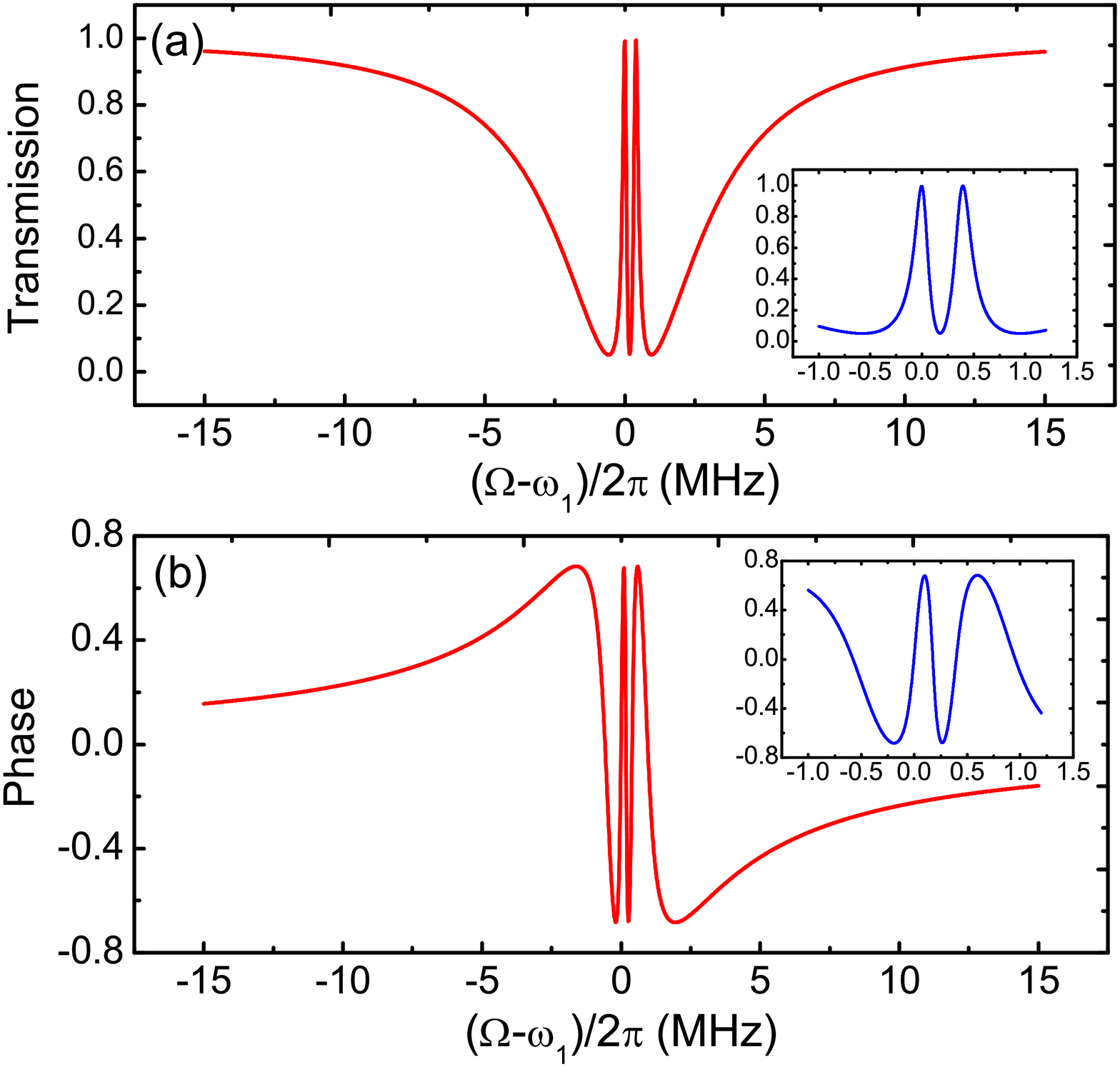}
\caption{(a) The magnitude and (b) phase of the probe transmission versus $(\Omega-\omega_1)/2\pi$ when the resonance frequencies of the two mechanical oscillators are different. The insets are the enlarged images in the middle of the figure. The pump field is red tuned by an amount of $\omega_m=(\omega_1+\omega_2)/2$ with $P_{pu}=1$ $\mu$W. Other parameters are $\omega_c=2\pi\times6.986$ GHz, $\kappa=2\pi\times6.2$ MHz, $\kappa_{e}=2\pi\times4.8$ MHz, $\omega_1=2\pi\times32.1$ MHz, $\omega_2=2\pi\times32.5$ MHz, $\gamma_1=\gamma_2=\gamma_m=2\pi\times930$ Hz, $g_1=2\pi\times39$ Hz, $g_2=2\pi\times44$ Hz.}
\end{figure} Note that the width of the transparency window when the two mechanical oscillators have equal resonance frequency ($\omega_1=\omega_2$) is larger than the case in the absence of the second mechanical oscillator ($g_2=0$). The reason is that the width of the transmission window is given by the modified mechanical damping rate $\gamma_1(1+2C_1)$ and $\gamma_1(1+C_1)$, respectively, where $C_1=4g_1^2n_p/\kappa\gamma_1$ denotes the optomechanical cooperativity \cite{Weis}.

In the following, we mainly consider the case in which the frequencies of the two mechanical oscillators are different but their damping rates are equal, which is the real situation in the experiment of Ref. \cite{Massel2}. When the pump field is red tuned by an amount $\omega_m=(\omega_1+\omega_2)/2$ with $P_{pu}=1\mu$W and the probe field is then scanned through the cavity resonance, in Fig. 3 we plot the magnitude and phase of the probe transmission as a function of the probe detuning $(\Omega-\omega_1)/2\pi$. Different from Fig. 2, there are two transmission peaks in the probe transmission spectrum, which displays that the input probe field could be simultaneously transparent at two symmetric frequencies. The inset of Fig. 3(a) is the zoom-in view of the two peaks, from which we can judge that the two peaks locate at $\Omega=\omega_1$ and $\Omega=\omega_2$, respectively. An intuitive physical picture explaining the bichromatic EMIT can be given in the energy level diagram shown in Fig. 1(b), where $n_p$ represents the number of the intracavity photon, and $m_1$ and $m_2$ denote the phonon numbers in the two mechanical oscillators with resonance frequencies $\omega_1$ and $\omega_2$, respectively. The pump field is red tuned by an amount $\omega_m$, i.e., $\Delta_{pu}=\omega_c-\omega_{pu}=\omega_m$, and the probe filed of frequency $\omega_{pr}=\omega_{pu}+\Omega$ scans through the cavity resonance. The simultaneous presence of pump and probe fields induces a modulation at the beat frequency $\Omega$ of the radiation pressure force acting on the mechanical oscillators. \begin{figure}
\centering
\includegraphics[width=8cm]{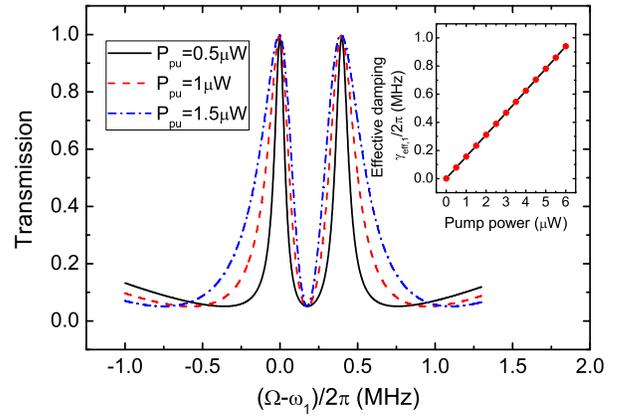}
\caption{The magnitude of the transmitted probe field as a function of $(\Omega-\omega_1)/2\pi$ with $\Delta_{pu}=\omega_m$ for $P_{pu}=0.5,$ 1, 1.5 $\mu$W, respectively. The inset plots the width of the transparency window at $\Omega=\omega_1$ as a function of the pump power.
Other parameters are the same as those in Fig. 3.}
\end{figure}When this modulation is close to the mechanical resonance frequency $\omega_k (k=1,2)$, the vibrational mode is excited coherently, giving rise to Stokes ($\omega_s=\omega_{pu}-\omega_k$) and anti-Stokes scattering ($\omega_{as}=\omega_{pu}+\omega_k$) of light from the strong pump field. If $\Delta_{pu}=\omega_{m}$, Stokes scattering is strongly suppressed because it is highly off-resonant with the microwave cavity and we can assume that only the anti-Stokes field builds within the cavity. However, when $\Omega=\omega_k$, the probe field is nearly resonant with the cavity, which is degenerate with the generated anti-Stokes fields. As a consequence, destructive interference between these two fields can suppress the build-up of an intracavity probe field and lead to two transparency windows in the probe transmission spectrum at $\Omega=\omega_1$ and $\Omega=\omega_2$. Fig. 3(b) shows that the phase of the probe transmission suffers a steep positive dispersion within the two transparency windows, which could be used as a tunable delay for microwave signals. Moreover, the width of the two transparency windows is given by $\gamma_{\mathrm{eff},k}=\gamma_{k}(1+C_k)$ with $C_k=4g_k^2n_p/\kappa\gamma_k(k=1,2)$.
The intracavity photon number $n_p$ can be greatly increased by enhancing the pump power $P_{pu}$ according to Eq. (12), which will result in broadening of the transparency windows, as shown in Fig. 4. From this figure, one can see that the transparency window is fully controllable via the applied microwave field, the window expanding and contracting with the power of the pump field. The inset of Fig. 4 shows a linear relationship between the pump power and the width of the window locating at $\Omega=\omega_1$.

\begin{figure}
\centering
\includegraphics[width=8cm]{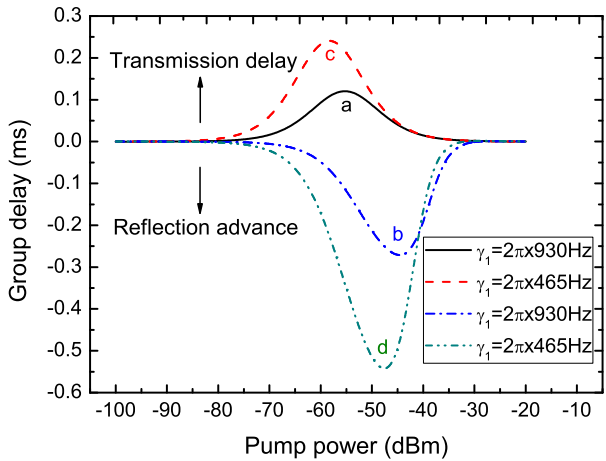}
\caption{Group delay $\tau_g$ of the transmitted (a,c) and reflected (b,d) probe field as a function of the pump power considering the effects of the damping rate of the mechanical oscillator. (a) and (b) are plotted using the experimental parameters. (c) and (d) are plotted assuming that the mechanical damping rate can be reduced. Other parameters are the same as those in Fig. 3.}
\end{figure}
Both the two transparency windows and the corresponding phase dispersion shown in Fig. 3 can be used for controlling the group delay of the microwave signals. For illustration, we consider the narrow transparency window and sharp phase dispersion at $\Omega=\omega_1$. The group delays of the transmitted and reflected probe fields as a function of the pump power are plotted in Fig. 5 with $\Delta_{pu}=\omega_m$ and $\Omega=\omega_1$. Solid curve (a) and dash curve (b) represent, respectively, the group delays of the transmitted and reflected probe filed with the realistic experimental parameter, i.e., $\gamma_1=2\pi\times930$ Hz. It can be seen that the group delays of the transmitted probe field are positive but those of the reflected probe field are negative when the power of the pump field is varied, which represent the slowing and advancing of the microwave signals, respectively. The maximum transmission group delay is $\tau_g^{(\mathrm{T})}\approx0.12$ ms and the reflection signal advance is $\tau_g^{(\mathrm{R})}\approx0.27$ ms. Moreover, we consider the effect of the mechanical damping rate on the group delay. In Ref. \cite{Zhou}, Zhou \emph{et al.} has shown analytically that the maximum group delay is inversely proportional to the mechanical damping rate. Here, we assume that $\gamma_1=2\pi\times465$ Hz, as shown in curve (c) and (d) in Fig. 5. One can see that the maximum delays are almost twice the values in curve (a) and (b). Therefore, the group delay of the microwave signal can be further increased by reducing the mechanical damping rates, which can be as low as several Hz \cite{Zhou,Andrews}. Compared to the generic circuit electromechanical systems in which an electromagnetic cavity is coupled to a single mechanical oscillator, one advantage of the 3-mode hybrid system we study here is that it allows for more controllability to realize double EMIT, which can also exist when the cavity-pump detuning $\Delta_{pu}=\omega_1$ or $\Delta_{pu}=\omega_2$. The two transparency windows still locate at $\Omega=\omega_1$ and $\Omega=\omega_2$. As a result, group delay of the probe field at a wider range of frequencies can be effectively tuned with negligible losses by the frequency and power of the pump field.

\section{conclusion}
To summarize, we have studied the tunable slowing and advancing of microwave signals based on double EMIT in a 3-mode circuit electromechanical system. When the frequencies of the two mechanical oscillators are different, two transparency windows appear in the probe transmission spectrum due to destructive interference between the probe field and the anti-Stokes filed caused by the two distinct mechanical vibrations. The narrow transparency window and the corresponding steep phase dispersion allow for controlling the group delay of the microwave probe field with negligible losses, which can be tuned by the power of the pump field. Our theoretical results show that the maximum group delay of the transmitted probe field is $\sim0.12$ ms, and the signal advance of the reflected probe field is about 0.27 ms for the experimental parameters in Ref. \cite{Zhou}.

\section*{ACKNOWLEDGMENTS} The authors gratefully acknowledge support
from National Natural Science Foundation of China (Grant No. 11304110), Jiangsu Natural Science Foundation (Grant No. BK20130413), and Natural Science Foundation of the Jiangsu Higher Education Institutions of China (Grant No. 13KJB140002).

\end{document}